\documentstyle[preprint,tighten,prd,aps,psfig,floats]{revtex}

\def\omb{\Omega_{\rm B}}
\def\ombh2{\Omega_{\rm B}h^2}
\def\om0{\Omega_0}
\def\oml{\Omega_\Lambda}
\def\del2{\Delta^2}
\def\delhh{\delta_{\rm H}}
\def\delh2{\delta_{\rm H}^2}
\def\sig8{\sigma_8}
\def\lya{Ly$\,\alpha$}
\def\ntw{\widetilde{n}}
\def\beq{\begin{equation}}
\def\eeq{\end{equation}}
\def\etal{{\it et~al.}}
\def\ie{{\it i.e.}\ }
\def\eg{{\it e.g.}\ }
\def\eq{Eq.\,}
\def\eqs{Eqs.\,}

\begin{document}
\preprint{UBC-COS-98-05}
\bibliographystyle{prsty}
\draft
\title{Limits on the gravity wave contribution to microwave anisotropies}
\author{J. P. Zibin and Douglas Scott}
\address{Department of Physics and Astronomy, 
University of British Columbia, 6224 Agricultural Road, 
Vancouver, BC V6T 1Z1 Canada}
\author{Martin White}
\address{Departments of Astronomy and Physics, University of Illinois at
Urbana-Champaign, 1110 West Green Street, Urbana, IL 61801-3080}
%\date{\today}
\maketitle

\begin{abstract}

\noindent We present limits on the fraction of large angle microwave
anisotropies which could come from tensor perturbations.  
We use the {\sl COBE\/} results as well as smaller scale CMB 
observations, measurements of galaxy correlations, abundances of galaxy
clusters, and \lya\ absorption cloud statistics.
Our aim is to provide conservative limits
on the tensor-to-scalar ratio for standard inflationary models.  
For power-law inflation, for example, we find $T/S<0.52$ at 
$95\%$ confidence, with a similar constraint for $\phi^p$ potentials.
However, for models with tensor amplitude 
unrelated to the scalar spectral index it is still currently 
possible to have $T/S>1$.

\end{abstract}
\pacs{PACS numbers: 98.80.-k, 98.70.Vc.}

\section{Introduction}

The presence of a primordial gravitational wave perturbation spectrum was an
early prediction of inflationary models of the big bang \cite{star}.
However, it was not until the results of the {\sl COBE\/} satellite mission
that it became possible to begin to meaningfully constrain the tensor
contribution to the overall perturbation spectrum \cite{kw,dhsst,ll,ss,sal}.  
In an early result, Salopek\cite{sal} found that, assuming power-law
inflation, tensors must contribute less than about $50\%$ of the cosmic
microwave background (CMB) fluctuations at the $10^\circ$ scale.

Since that time, ground- and balloon-based experiments have begun to fill in
the smaller-scale regions of the CMB power spectrum.
These scales are crucial for
constraining the gravity wave contribution because the tensor spectrum is
expected to be negligible on scales finer than $\sim1^\circ$, and therefore
large-scale power greater than that expected for scalars can be attributed to 
tensors.  Markevich and Starobinsky \cite{ms} have set some
stringent limits on the tensor
contribution.  For example, they found that the ratio of tensor to scalar
components of the CMB spectrum is $T/S < 0.7$ at $97.5\%$ confidence for
a flat, cosmological-constant-free universe with
$H_0 = 50\:{\rm kms^{-1}Mpc^{-1}}$.
However, their analysis used a limited CMB data set and considered only a
restricted set of values for the cosmological parameters.  Recently
Tegmark \cite{teg} has performed an analysis using a compilation of CMB data,
and found a $68\%$ upper confidence limit of 0.56 on the tensor to scalar
ratio.  In this work specific inflationary models were not considered,
but a number of parameters were allowed to vary freely.
In another recent study
Lesgourgues \etal \cite{lps} analysed a particular broken-scale-invariance
model of inflation with a steplike primordial perturbation spectrum, and found
that the tensor to scalar ratio can reach unity.
Melchiorri \etal \cite{mssv} placed limits on tensors allowing for a blue 
scalar spectral index, and indeed found that blue spectra and a large
tensor component are most consistent with CMB observations.

Our aim here is to provide a more comprehensive answer (or set of answers) to 
the question: how big can $T/S$ be?
We present constraints on tensors for specific models of inflation as
well as for freely varying parameters.  In all cases we marginalize over the
important, but as yet undetermined, cosmological parameters.
We use both {\sl COBE\/} and small-scale CMB data, as well as information
about the matter power spectrum from galaxy correlation, cluster abundance,
and Lyman $\alpha$ forest measurements.
We refer to these various measurements of the power spectra as ``data sets''.
We additionally consider the effect,
for each data set, of various observational
constraints on the cosmological parameters, such as the age of the universe,
cluster baryon density, and recent supernova measurements.
We refer to these constraints as ``parameter constraints''
(this separation between ``data sets'' and ``parameter constraints'' is
somewhat subjective, but dealt with consistently in our Bayesian approach;
it is conceptually simpler to consider power spectrum constraints as
measurements with some Gaussian error, while regarding
allowed limits on cosmological parameters as restrictions on parameter space).
Finally we consider what implications our results have for the direct detection
of primordial gravity waves.

\section{Inflation models}

Our goal is to provide limits on the tensor contribution to the primordial
perturbation spectra using a variety of recent observations.  In models of
inflation, the scalar (density) and tensor (gravity wave) metric perturbations
produced during inflation are specified by two spectral functions,
$A_{\rm S}(k)$ and $A_{\rm T}(k)$, for wave number $k$.
These spectra are determined by the inflaton potential $V(\phi)$ and its
derivatives \cite{InfReview}.
However, when comparing model predictions with actual observations of the
CMB, it is more useful to translate the inflationary spectra into the
predicted multipole expansions of the CMB temperature field: 
$\Delta T/T(\theta,\phi) = \sum_{\ell m}a_{\ell m}Y_{\ell m}(\theta,\phi)$, 
where $Y_{\ell m}(\theta,\phi)$ are the spherical harmonics.  
The spectrum $C_\ell \equiv \langle|a_{\ell m}|^2\rangle$ can be decomposed
into scalar and tensor parts, $C_\ell=C_\ell^{\rm S}+C_\ell^{\rm T}$.
In the literature the tensor to scalar ratio is conventionally specified 
either at $\ell=2$ or in the spectral plateau at $\ell\simeq 10-20$.
Here we have chosen the $\ell=2$ or quadrupole moments of the temperature
field, and write $S\equiv 5C_2^{\rm S}/4\pi$ and $T\equiv 5C_2^{\rm T}/4\pi$
as usual \cite{tw}.

In order to constrain the tensor contribution $T/S$, we need to specify the
particular model of inflation under consideration.  This is because the model
may provide a specific relationship between the ratio $T/S$ and the scalar
spectral index $n_{\rm S}$.
Except in Sec.~\ref{sec:open} we only consider spatially flat inflation models 
(\ie $\om0+\oml=1$, where $\om0$ and $\oml=\Lambda/(3H_0^2)$ are the fractions
of critical density due to matter and a cosmological constant, respectively).
In addition, we do not consider the ``quintessence'' models
\cite{quint1,quint2}, where a significant fraction of the critical density is
currently in the form of a scalar field with equation-of-state different from
that of matter, radiation, or cosmological constant (although it would not
be difficult to extend our results for explicit models with recent epoch
dynamical fields).
We will also restrict ourselves to models which
use the slow-roll approximation, and incorporate only a single 
{\it dynamical\/} field -- a class of models sometimes called
``chaotic inflation'' \cite{chaotic}.
This is not as restrictive as it might sound, since most viable inflationary
models are of this form.
Although some genuinely two-field models are known \cite{multifield},
many multi-field models, the ``hybrid'' class, have only one field dynamically
important and in these cases we effectively regain the single field case
\cite{hybrid,lid97}.
In addition, theories which modify general relativity (\eg ``extended''
inflation \cite{lastein,kolb}) can often be recast as ordinary general
relativity with a single effective scalar field \cite{conformal,lid97}.

It is convenient to classify inflationary models as either ``small-field'',
``large-field'', or the already mentioned hybrid models \cite{dkk}.
Small-field models are characterized by an inflaton field which rolls from
a potential maximum towards a minimum at $\langle\phi\rangle \neq 0$.  
These models generally produce negligible tensor contribution, but may result
in the spectral index $n_{\rm S}$ differing significantly from scale invariance
\cite{InfReview}.
In hybrid models, the important scalar field rolls towards a potential minimum
with non-zero vacuum energy.  These models also typically have very small
$T/S$, and the scalar index can be greater than unity \cite{InfReview}.
The large-field models involve so-called ``chaotic'' initial conditions, where 
an inflaton initially displaced from the potential minimum rolls towards the
origin.  Large-field models can produce large $T/S$ and $1-n_{\rm S}$,
and these are the models considered in this paper.
This is not to say that small-field and hybrid models are not interesting;
on the contrary, current views of inflation in the particle physics context
suggest that $T/S$ is expected to be small \cite{lyth97}.
However, large-field models must be considered when examining the
observational evidence for a large tensor contribution.

It is also worth pointing out that we could construct models with a dip in
the scalar power spectrum at large scales which compensates for the tensor
contribution.  Although we have not explored detailed models, we imagine that
in principle models could be constructed with arbitrarily high $T/S$.
We consider all such models with features at relevant scales to be unappealing
unless there are separate physical arguments for them.

In addition to considering models with free scalar index and tensor
contribution $T/S$, we shall thus focus on two classes of inflationary models
which can be considered representative of those predicting large gravity wave
contributions.  Both are restricted to ``red'' spectral tilts, 
$n_{\rm S}\leq 1$.  The first, ``power-law inflation'' (PLI) \cite{aw,lm}, is 
characterized by exponential inflaton potentials of the form
\beq
  V(\phi) \propto \exp\sqrt{\frac{16\pi\phi^2}{qm_{\rm Pl}^2}},
\eeq
and results in a scale factor growth $a(t)\propto t^q$, hence the name.
For PLI the tensor-to-scalar ratio in $k$-space can be calculated exactly
as a function of $n_{\rm S}$:
\beq
  { A_{\rm T}^2(k)\over A_{\rm S}^2(k) }
  = {1-n_{\rm S}\over 3-n_{\rm S}} \qquad .
\eeq
Note the tensor contribution is directly related to the scalar spectral
index $n_{\rm S}$, which is further related to the tensor spectral index
$n_{\rm T}=n_{\rm S}-1$ in this model.
Converting from $k$-space to the observed anisotropy spectrum introduces a
dependence on the cosmological constant which can be approximated by \cite{tw}
\beq
  T/S = -7\ntw\left[0.97+0.58\ntw+0.25\oml
      -\left(1+1.1\ntw+0.28\ntw^2\right)\oml^2\right],
\label{ts}
\eeq
where $\ntw \equiv n_{\rm S}-1 = n_{\rm T}$.
The dependence on $\Lambda$ arises because of different evolution for scalars
and tensors when $\Lambda$ dominates at late times.  
The dependence on other cosmological parameters is negligible \cite{tw}.

We also consider the large-field polynomial potentials,
\beq
  V(\phi) \propto \phi^p,
\eeq
for integral $p>1$ \cite{linde}.  In this case both $n_{\rm S}$ and 
$n_{\rm T}$ are determined by the exponent $p$ \cite{InfReview}:
\beq
  n_{\rm S}=1-\frac{2p+4}{p+200}, \label{np}
\eeq
\beq
  n_{\rm T}=\frac{-2p}{p+200}. \label{ntp}
\eeq
The tensor index may be related to $T/S$ through the consistency 
relation \cite{tw}
\beq
 \frac{T}{S} = -7\frac{f_{\rm T}^{(0)}}{f_{\rm S}^{(0)}} n_{\rm T}, \label{tsp}
\eeq
where the cosmological parameter dependence, again dominated 
by $\oml$, can be approximated by
\begin{eqnarray}
  f_{\rm S}^{(0)} &=& 1.04-0.82\oml+2\oml^2, \\
  f_{\rm T}^{(0)} &=& 1.0-0.03\oml-0.1\oml^2.
\end{eqnarray}

As a third possibility, we will also consider models with scalar
index varying over the range $n_{\rm S}=0.8 - 1.2$, but with an 
independently varying tensor contribution $T/S$.

\section{Microwave background anisotropies} \label{cmb}

In order to evaluate likelihoods and confidence limits for $T/S$ based on
CMB measurements, we performed $\chi^2$ fits of model $C_\ell$ spectra to
CMB data.  We did this for a set of ``band-power'' estimates of anisotropy
at different scales, and separately for the {\sl COBE\/} data themselves.
For our first approach, we used a collection of binned data to represent
the anisotropies as a function of $\ell$.
Specifically we took the flat-spectrum effective quadrupole values listed in
Smoot and Scott \cite{rpp} and binned them into nine intervals separated
logarithmically in $\ell$.
We chose this simplified approach since we anticipated a large computational
effort in covering a reasonably large parameter space.
The use of binned data has been shown elsewhere \cite{BondJK}
to give similar results to more thorough methods.  If anything, 
there is a bias towards lowering the height of any acoustic peak,
inherent in the simplifying assumption of symmetric Gaussian error
bars \cite{BondJK2}; for placing upper limits on $T/S$ our approach
is therefore conservative.  We are also erring on the side of caution 
by using the binned data only up to the first acoustic peak, 
neglecting constraints from detections and upper limits at 
smaller angular scales.

We ignored the effect of reionization on the $C_\ell$ spectra.  Reionization 
to optical depth $\tau$ reduces the power of small-scale anisotropies 
by $e^{-2\tau}$.  Thus, in placing upper limits on $T/S$, it is 
conservative to set $\tau=0$.

A fitting function for the spectrum, valid up to the first peak at
$l \simeq 220$, has been provided by White \cite{w}:
\beq
  C_l(\nu) = \left(\frac{l}{10}\right)^\nu C_l(\nu=0),\label{nuspec}
\eeq
where $\nu$ is the (nearly) degenerate combination of cosmological parameters
\beq
  \nu \equiv n_{\rm S}-1-0.32\ln(1+0.76r)+6.8(\ombh2-0.0125)
           -0.37\ln(2h)-0.16\ln(\om0).  \label{nu}
\eeq
Here $r \equiv 1.4C_{10}^{\rm T}/C_{10}^{\rm S}$ is the tensor to scalar
ratio at $\ell=10$, normalized to provide $r=T/S$ for $\om0=1$ and
$n_{\rm S}\rightarrow 1$.  The parameter $h$ is defined through
$H_0=100h\:{\rm kms^{-1}Mpc^{-1}}$, and $\omb$ is the fraction of the critical
density in baryons.  Thus the standard CDM (sCDM) spectrum 
is specified by $\nu=0$.
We found that the $\oml$ dependence of $r$ can be well captured by introducing
the rescaled variable $r^\prime$, defined by
\beq
  r^\prime = {r\over 0.94+1.105\oml^{3.75}},
\eeq
and setting $r^\prime\equiv T/S$.

We fitted the model spectra of \eq(\ref{nuspec}) to the binned data as
follows.  For each combination of parameters ($h,\ombh2,\om0,n_{\rm S},T/S$) we
normalized the model spectrum to the binned data, and evaluated the
likelihood ${\cal L}(h,\ombh2,\om0,n_{S},T/S) \propto \exp(-\chi^2/2)$.
Next this likelihood was integrated, uniformly in the parameter, over the
ranges of $h=0.5 - 0.8$, $\ombh2=0.007 - 0.024$, and $\om0=0.25 - 1$, subject
to the constraints of \eq(\ref{ts}) for PLI and \eqs(\ref{np}), (\ref{ntp}), 
and (\ref{tsp}) for polynomial potentials.  For the case of free $T/S$, 
the scalar index was varied in the range $n_{\rm S}=0.8 - 1.2$.
Finally the resultant ${\cal L}(T/S)$ was normalized to a peak value of unity
and the $95\%$ confidence limits evaluated.  We tried to choose reasonable
ranges for the prior probability distributions of the ``nuisance parameters'',
guided by the current weight of evidence.  We checked that mild departures
from our adopted ranges lead to only small modifications to our results.
However, we caution that our conclusions will not necessarily be applicable
for models which lie significantly outside the parameter space we considered.
In addition, note that according to \eq(\ref{nu}) we can crudely estimate 
an upper limit on $T/S$ by combining the observational lower limit on $\nu$ 
with the maximal baryon density and minimal $\om0$ and $h$ from our parameter 
ranges.  However, this 
turns out to be an overly conservative estimate:  for example, for 
$n_{\rm S}=1$, and using a lower limit of $\nu=-0.2$, \eq(\ref{nu}) gives 
an upper limit of $T/S=3.8$, compared with the limit $T/S=1.6$ from 
Sec.~\ref{results}.  

For the separate constraint from the {\sl COBE\/} data,
we used the software package
CMBFAST \cite{sz} to calculate likelihoods based only on the {\sl COBE\/}
results at large scales.  CMBFAST calculates the spectrum using a line-of-sight
integration technique.  It then calculates likelihoods by finding a quadratic
approximation to the large scale spectrum and using the {\sl COBE\/} fits of
Bunn and White \cite{bw}.  
These likelihoods were integrated and $95\%$ limits calculated as above,
except that the baryon density was fixed at $\omb=0.05$ to save computation
time (and since $\omb$ has negligible affect at these scales).
The results of this procedure are presented in Sec.~\ref{results}.

\section{Large-scale structure}

\subsection{Galaxy correlations}

We next applied observations of galaxy correlations to constrain $T/S$
indirectly through the power spectrum of the density fluctuations,
$\del2(k)$. The power spectrum $\del2(k)$ is expressed, following
Bunn and White \cite{bw}, by
\beq
\del2(k) = \delh2 \left(\frac{ck}{H_0}\right)^{3+n_{\rm S}} T^2(k). \label{ps1}
\eeq
Here $\delhh$ is the ($\om0$, $n_{\rm S}$, and $r$ dependent) normalization 
described in Sec.~\ref{clustab}, and $T(k)$ is the transfer function which 
describes the evolution of the spectrum from its primordial form to the 
present.

We explicitly used for the transfer function the fit of
Bardeen \etal \cite{bbks},
\beq
T(q)=\frac{\ln(1+2.34q)}{2.34q} 
     \left[1+3.89q+(16.1q)^2+(5.46q)^3+(6.71q)^4\right]^{-1/4},
\label{ps2}
\eeq
with the scaling of Sugiyama \cite{s}
\beq
q = \frac {k(T_{\gamma0}/2.7 \: {\rm K})^2}
    {\om0 h^2 \exp\left(-\omb-\sqrt{h/0.5}\:\omb/\om0\right)}. \label{ps3}
\eeq
Here $T_{\gamma0}$ is the temperature of the CMB radiation today.

We performed $\chi^2$ fits of the (unnormalized) model power spectrum given
by \eqs(\ref{ps1})~-~(\ref{ps3}) to the compilation of data provided in
Table~I of Peacock and Dodds \cite{pd}, excluding their four smallest
scale data points.
These points were omitted because, while there are theoretical reasons
\cite{sw} to expect that the galaxy bias approaches a constant on large
scales, at the smallest scales the assumption of a linear bias appears
to break down \cite{peacock}.  
Here we are fitting for the {\it shape\/} of the matter power spectrum,
ignoring the overall amplitude, since the normalization is complicated by
the ambiguities of galaxy biasing.  

The fitting was performed in exactly the same way as was described in
Sec.~\ref{cmb} for the binned microwave anisotropies.
Namely the model curves were normalized to the Peacock and Dodds data,
the integrated likelihood was calculated, and the $95\%$ confidence limits
for $n_{\rm S}$ were evaluated.  Since the shape of the power spectrum
[\eq(\ref{ps3})] is independent of the tensor amplitude, this technique can
only provide limits on $T/S$ when $T/S$ is determined by the spectral index.
That is, the galaxy correlation data can only constrain $T/S$ for our
PLI and $\phi^p$ cases, using the relationships [\eq(\ref{ts}) or
\eqs(\ref{np}), (\ref{ntp}), and (\ref{tsp})] between $T/S$ and $n_{\rm S}$.

\subsection{Cluster abundance}
\label{clustab}

A very useful quantity for constraining the {\it amplitude\/} of the power
spectrum is the dispersion of the density field smoothed on a scale $R$,
defined by
\beq
  \sigma^2(R) = \int_0^\infty W^2(kR) \del2(k) \frac{dk}{k}. \label{sigdef}
\eeq
Here $W(kR)$ is the smoothing function, which we take to be a 
spherical top-hat specified by
\beq
W(kR) = 3\left[\frac{\sin(kR)}{(kR)^3} - \frac{\cos(kR)}{(kR)^2} \right].
\eeq
Traditionally the dispersion is quoted at the scale $8h^{-1}$ Mpc, 
and given the symbol $\sig8$.  For our experimental value we used
the result of Viana and Liddle \cite{vl98}, who analysed the abundance
of large galaxy clusters to obtain
\beq
\sig8 = 0.56\, \om0^{-0.47} \label{sig8},
\eeq
with relative $95\%$ confidence limits of $-18\, \om0^{0.2\log_{10}\om0}$ 
and $+20\, \om0^{0.2\log_{10}\om0}$ percent.  Several other estimates 
have been published; the one we used is fairly representative, 
and with a more conservative error bar than most.

To compare this experimental result with the model value predicted by
\eq(\ref{sigdef}), we must fix the normalization $\delhh$.  We used
the result of Liddle \etal \cite{llvw} who fitted $\delhh$ using the
{\sl COBE\/}
large scale normalization to obtain
\beq
10^5\: \delhh (n_{\rm S},\om0) = 1.94\: \om0^{-0.785-0.05\ln\om0}
   \exp[f(n_{\rm S})],
\eeq
where
\beq
f(n_{\rm S})=\left\{
\begin{array}{ll}
-0.95\ntw-0.169\ntw^2,\;\; & \mbox{No tensors,} \\
 1.00\ntw+1.97 \ntw^2,     & \mbox{PLI.}
\end{array}
\right.
\eeq
For the case of non-PLI tensors, we used 
the fitting form of Bunn \etal 
\cite{blw}:
\beq
10^5\: \delhh = 1.91\: \om0^{-0.80-0.05\ln\om0} 
\frac{\exp(-1.01\ntw)}{\sqrt{1+\left(0.75-0.13\oml^2\right)r}}
\left(1+0.18\ntw\oml-0.03r\oml\right).
\eeq

We calculated likelihoods for our model $\sig8$ using
a Gaussian with peak and $95\%$ limits specified by
\eq(\ref{sig8}), and then integrated ${\cal L}$ and found limits 
for $T/S$ as in the binned microwave case.

\subsection{Lyman $\alpha$ absorption cloud statistics}

Another measure of the amplitude of the matter power spectrum 
has been obtained recently by Croft \etal \cite{cwphk}, who 
analysed the Lyman $\alpha$ (\lya) absorption forest in the spectra
of quasars at redshifts $z\simeq2.5$.  These 
results apply at smaller comoving scales than the cluster abundance 
$\sig8$ measurements, and hence are potentially more 
constraining.  Croft \etal\ found 
\beq
\del2(k_p) = 0.57\, ^{+0.26}_{-0.18}
\eeq
at $1\sigma$ confidence, where the effective wavenumber
$k_p=0.008(\mbox{km s}^{-1})^{-1}$ at $z=2.5$.

These results cannot be directly compared with the model 
predictions of \eq (\ref{ps1}), because \eq (\ref{ps1}) 
provides its predictions for the current time, \ie $z=0$.  
To translate to $z=2.5$, we must first convert the model 
$k$ from the comoving $\mbox{Mpc}^{-1}$ units conventionally 
used in discussions of the matter power spectrum to 
$(\mbox{km s}^{-1})^{-1}$ at $z=2.5$, using 
\beq
k[(\mbox{km s}^{-1})^{-1}] = \frac{1+z}{H(z)} \: k[\mbox{Mpc}^{-1}],
\eeq
where
\beq
H(z) = H_0\sqrt{\om0(1+z)^3 + \oml}
\eeq
for flat universes.

Next, we must consider the growth of the perturbations 
themselves.  In a critical density universe (and assuming linear theory),
the growth law is simply $\del2(k,z) = \del2(k,0)(1+z)^{-2}$.  As $\oml$ 
increases, the growth is suppressed, and this can be 
accounted for by writing 
\beq
\del2(k,z) = \del2(k,0) \frac{g^2[\Omega(z)]}{g^2(\om0)} 
                        \frac{1}{(1+z)^2},
\eeq
where the growth suppression factor $g(\Omega)$ can be 
accurately parametrized by \cite{cpt} 
\beq
g(\Omega) = \frac{5}{2}\Omega\left(\frac{1}{70} + \frac{209\Omega}{140}
          - \frac{\Omega^2}{140} + \Omega^{4/7}\right)^{-1}\!,
\eeq
and the redshift dependence of $\Omega$ is given by 
\beq
\Omega(z) = \om0\ \frac{(1+z)^3}{1-\om0+(1+z)^3\om0},
\eeq
all for spatially flat universes.

We calculated likelihoods using the normalized model predictions of
\eq (\ref{ps1}), translated to $z=2.5$ as described above, and then obtained
limits for $T/S$ as in the cluster abundance case.

\section{Parameter constraints}

\subsection{Age of the universe}

In flat $\Lambda$ models, the age of the universe is \cite{cpt}
\beq
  t_0=\frac{2}{3H_0}\;\frac{\sinh^{-1}\left(\sqrt{\oml/\om0}\right)}
                           {\sqrt{\oml}}.
\label{age}
\eeq
During the integration of the likelihoods,
we investigated the effect of imposing a constraint on 
the parameters $h$ and $\om0$, so that regions of parameter space 
corresponding to ages below various limits were excluded.  This simply
corresponds to a more complex form for the priors on the 
parameters.  The precise limit on the age of the universe 
is a matter of on-going debate (\eg \cite{vsb,chab}).  
A lower limit of around $11$ Gyr now seems to be the norm, 
so we considered this case explicitly.  We also considered 
the effect of a more constraining limit of $13$ Gyr, still 
preferred by some authors.

\subsection{Baryons in clusters}

Recent measurements of the baryon density in clusters have suggested low
$\om0$ for consistency with nucleosythesis.  We chose to use the results
of White and Fabian \cite{wf} for the baryon density
\beq
  \frac{\omb}{\om0} = (0.056\pm 0.014) h^{-3/2},
\eeq
where the errors are at the $1\sigma$ level.
We explored the implications of applying this constraint
during the likelihood integrations, by adding a term
\beq
  \left(\frac{h^{3/2}\omb/\om0 - 0.056}{0.014}\right)^2
\eeq
to each value of $\chi^2$.

\subsection{Supernova constraints}

Measurements of high-$z$ Type-Ia supernovae (SNe Ia) are in principle
well-suited to constraining $\om0$ on the assumption of a flat $\Lambda$
universe, since such measurements are sensitive to (roughly) the difference
between $\om0$ and $\oml$.
We used the experimental results of Filippenko and Riess of the High-$z$
Supernova Search team \cite{fr}, who found for flat $\Lambda$ models
\beq
  \om0 = 0.25 \pm 0.15
\eeq
at $1\sigma$ confidence.  We also investigated the effect of
applying this constraint as above.

\section{Open models} \label{sec:open}

For models with open geometry the situation is more complicated, and so
we restrict ourselves to a brief discussion here.
In addition
to the added technical complexity involved in working in hyperbolic spaces,
the presence of an additional scale, the curvature scale, renders ambiguous
the meaning of scale-invariant fluctuations.
For the most obvious scale-invariant spectrum of gravity wave modes, the
quadrupole anisotropy actually diverges!
For this reason one requires a definite calculation of the fluctuation
spectrum from a well realized open model.
The advent of open inflationary models \cite{OpenInf} has allowed, for the
first time, a {\it calculation\/} of the spectrum of primordial fluctuations
in an open universe.
As with all inflationary models, a nearly scale-invariant spectrum of
gravitational waves (tensor modes) is produced \cite{TanSas,BucCoh}.
The size of these modes in $k$-space, and their relation to the spectral
index, is not dissimilar to the flat space models we have been considering.
In the inflationary open universe models the spectrum of perturbations is
cut-off at large spatial scales, leading to a finite gravity wave spectrum.
However, the exact scale of the cutoff depends on details of the model,
introducing further model dependence into the $\ell$-space predictions.

\begin{figure}
\centerline{\psfig{figure=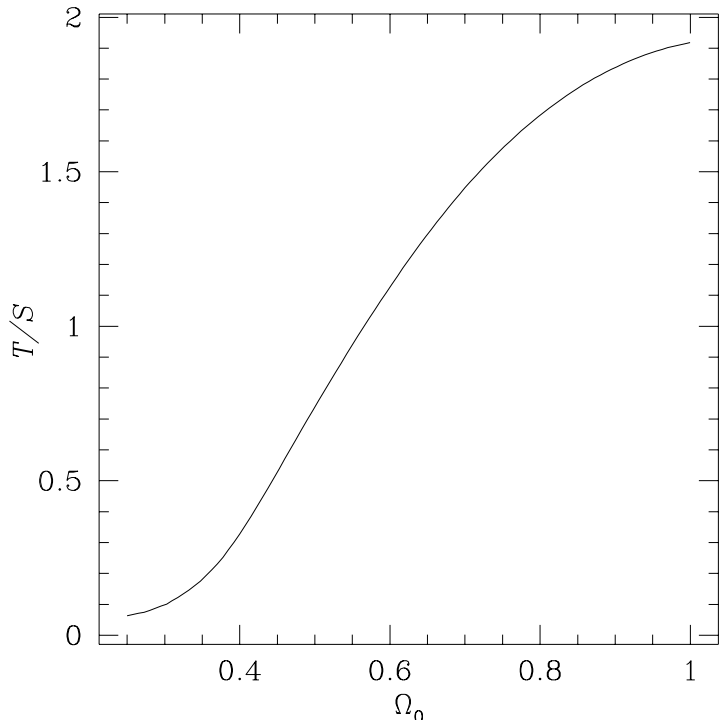,height=10cm}}
\caption{$95\%$ upper confidence limits on $T/S$ for the open models 
described in the text.  The cluster abundance data set was used with 
no parameter constraints.}
\label{openfig}
\end{figure}

Since gravitational waves provide anisotropies but no density fluctuations,
their presence will in general lower the normalization of the matter power
spectrum (for a fixed large angle CMB normalization).
Open models
already have quite a low normalization \cite{WhiSil,WhiSco}, so the most
conservative limits on gravity waves come from models which produce the
{\it minimal\/} tensor anisotropies, i.e.~where the cutoff operates as
efficiently as possible.  The {\sl COBE\/} normalization for such models
with PLI is \cite{OpenTensor}
\begin{equation}
10^5\: \delta_{\rm H}=1.95\: \Omega_0^{-0.35-0.19\ln\Omega_0+0.15\widetilde{n}}
  \exp\left( 1.02 \widetilde{n} + 1.70 \widetilde{n}^2 \right) \, .
\end{equation}
Combining this normalization with the cluster abundance gives a strong
constraint on $T/S$.  We show in Fig.~\ref{openfig}
the 95\% CL upper limit on $T/S$ as a function of $\om0$ in these models.

\section{Results} \label{results}

\begin{figure}
\centerline{\psfig{figure=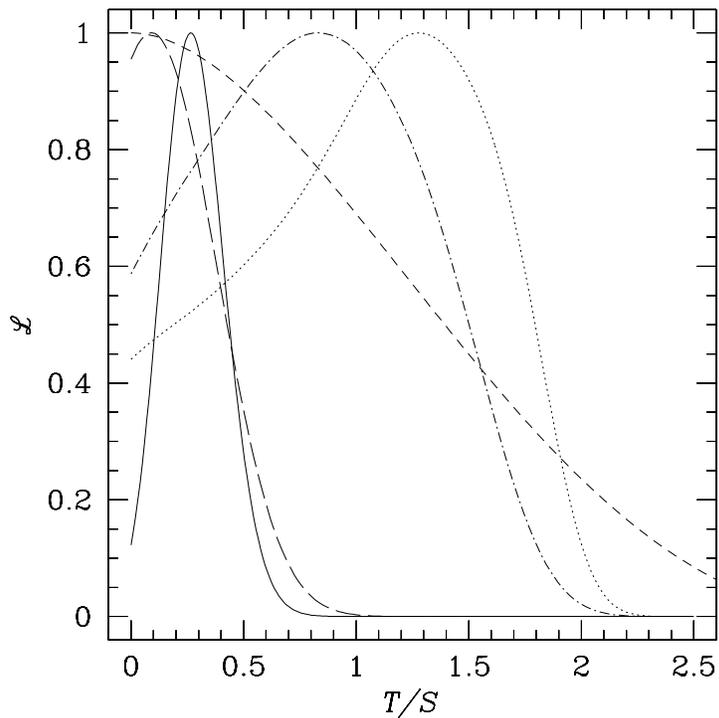,height=10cm}}
\caption{Integrated likelihoods versus $T/S$ for the various data sets:  
short-dashed, long-dashed, dotted, dash-dotted, and solid 
curves represent respectively {\sl COBE\/}, binned CMB, cluster abundance, 
\lya\ absorption, and combined data.  All curves are for PLI 
inflation, using the priors discussed in Sec.~\ref{cmb}, with no additional
parameter constraints.}
\label{noconstr}
\end{figure}

Figure \ref{noconstr} presents likelihoods, integrated over the 
parameter ranges described above, and plotted versus $T/S$, 
for the various data sets, and specifically for PLI models.  For 
the curve labelled ``combined'', likelihoods 
for each data set (except the {\sl COBE\/} data) were multiplied together 
before integration.  (Including the {\sl COBE\/} data would have 
been redundant, since the binned CMB set already contains the {\sl COBE\/}
results.)  Thus the ``combined'' values represent joint likelihoods  
for the relevant data sets, on the assumption of independent data.
Note that the combined data curve of Fig.~\ref{noconstr} differs
significantly from the product of the already marginalized curves 
for the different data sets, which indicates that parameter covariance 
is important here.  Also, the maximum joint likelihood in 
Fig.~\ref{noconstr} corresponds to $\chi^2 \simeq 9$, which indicates 
a good fit for the 15 degrees of freedom involved.

\begin{figure}
\centerline{\psfig{figure=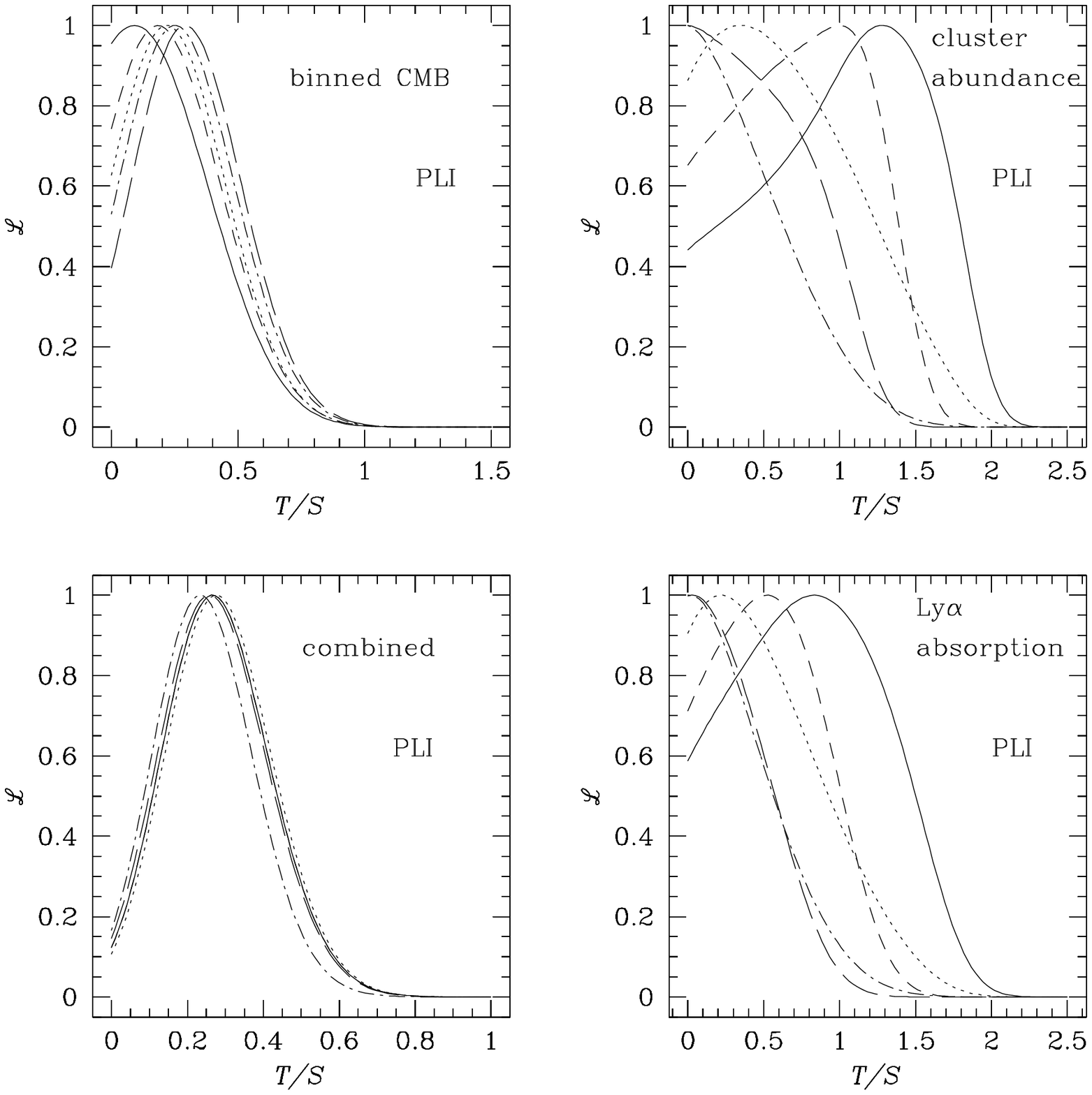,height=18cm}}
\caption{Integrated likelihoods versus $T/S$ for PLI and for the 
various data sets; clockwise from upper left: binned CMB, 
cluster abundance, \lya\ absorption, and combined data.  
Solid, short-dashed, long-dashed, 
dotted, and dot-dashed curves represent no constraint, $t_0>11$ Gyr, 
$t_0>13$ Gyr, cluster baryon fraction, and SNe Ia parameter constraints,
respectively.}
\label{plifig}
\end{figure}

Figure \ref{plifig} displays integrated 
likelihoods versus $T/S$ for each data set and for the combined 
data, again on the assumption of PLI.  
The effect of each parameter constraint is illustrated.  The 
{\sl COBE\/} data shape constraint is very weak, and exhibits essentially 
no cosmological parameter dependence, as expected.  Thus 
the parameter  constraints have little effect on the likelihoods, 
and the curves are not shown here.  Cluster 
abundance is not much more constraining than the {\sl COBE\/} shape, 
but exhibits considerably stronger cosmological parameter 
dependence, and hence is affected substantially by the 
various parameter constraints.  The matter power 
spectrum shape constraint is so weak that we do not plot it here.  
The strongest constraint comes from the binned CMB data, and 
indeed these data dominate the joint results.

We can understand the general features of the large parameter 
dependence exhibitted by the likelihoods for the matter spectrum 
data sets as follows.  Near sCDM parameter values, it is 
well known that the matter power spectrum contains too much 
small-scale power when {\sl COBE}-normalized at large scales.
The presence of tensors improves the fit at small scales by 
decreasing the {\it scalar} normalization at {\sl COBE\/} scales.
Reducing $h$ or $\om0$, however, also decreases the power at small 
scales, improving the fit over sCDM, and thus reducing the need 
for tensors.  When an age constraint is applied, we force the 
model towards lower $h$ and $\om0$ according to \eq (\ref{age}), and
hence towards lower $T/S$, as is seen in Fig.~\ref{plifig}.  The cluster 
baryon and supernova constraints similarly move us to smaller $\om0$.

\begin{figure}
\centerline{\psfig{figure=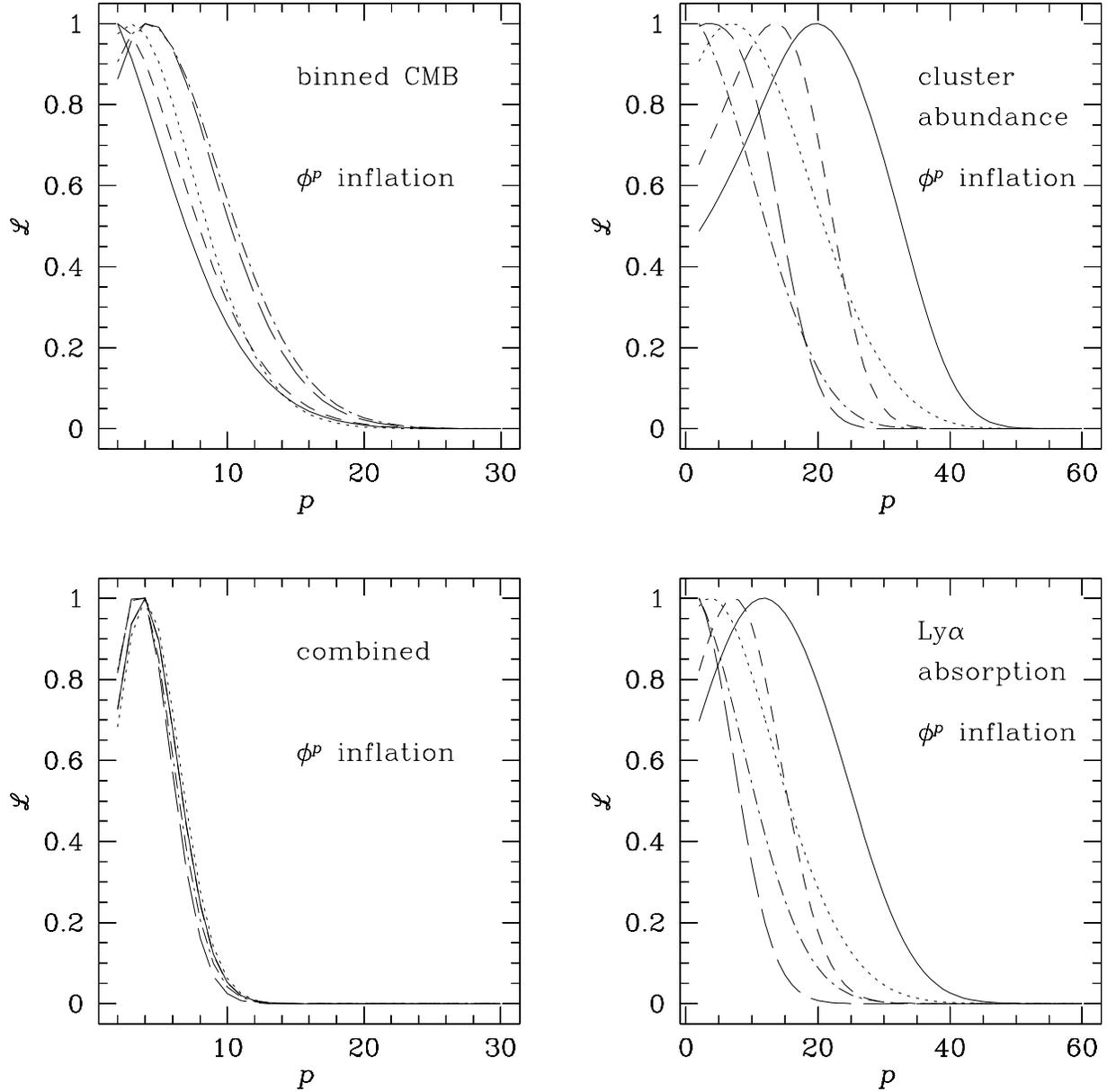,height=18cm}}
\caption{Integrated likelihoods versus $p$ for $\phi^p$ 
inflation and for various data sets;  
clockwise from upper left: binned CMB, cluster abundance, \lya\ absorption, 
and combined data.  Solid, short-dashed, long-dashed, dotted, and dot-dashed 
curves represent no constraint, $t_0>11$ Gyr, $t_0>13$ Gyr,
cluster baryon fraction, and SNe Ia parameter constraints, respectively.}
\label{phipfig}
\end{figure}

\begin{figure}
\centerline{\psfig{figure=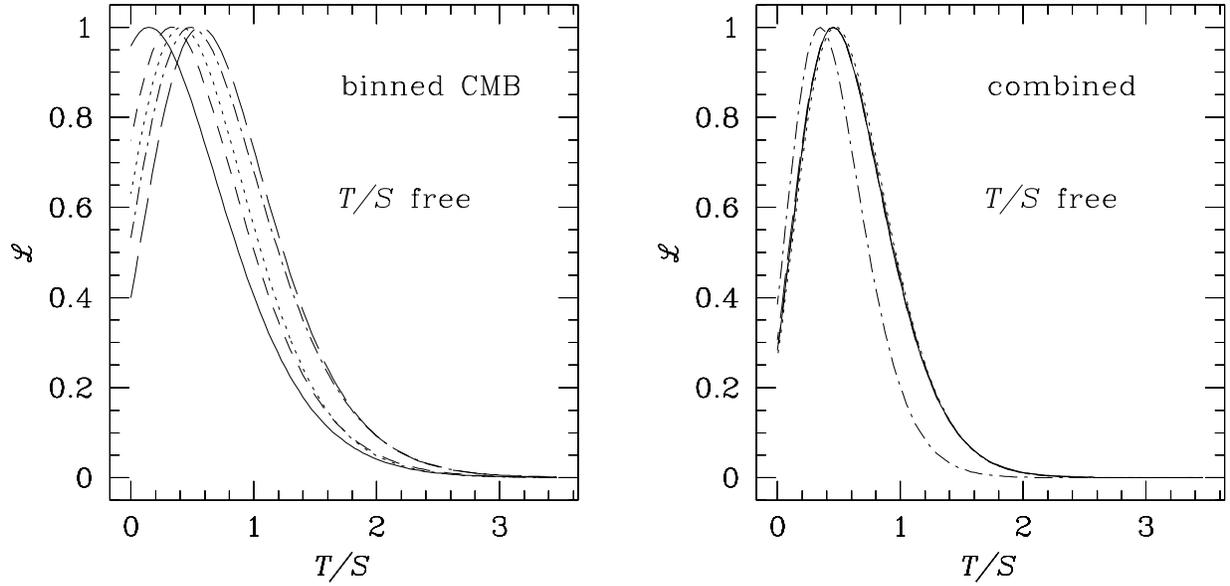,height=18cm}}
\caption{Integrated likelihoods versus $T/S$ for $T/S$ free and $n_{\rm S}=1$,
for binned CMB (left) and combined data (right).
Solid, short-dashed, long-dashed, 
dotted, and dot-dashed curves represent no constraint, $t_0>11$ Gyr, 
$t_0>13$ Gyr, cluster baryon fraction, and SNe Ia parameter constraints,
respectively.}
\label{freefig}
\end{figure}

Figure \ref{phipfig} displays likelihoods versus 
$p$ for $\phi^p$ inflation, while Fig.~\ref{freefig} 
presents likelihoods versus $T/S$ for the case of free 
tensor contribution and $n_{\rm S}=1$.  In all plots, curves have been
omitted for the very weakly constraining data sets.  The curves 
of Fig.~\ref{phipfig} closely resemble those of 
Fig.~\ref{plifig}.  This is because, for $p\gg2$, 
\eqs(\ref{np}), (\ref{ntp}), and (\ref{tsp}) 
give $T/S\simeq-6.85\,\ntw$, which is similar to the PLI result of 
\eq(\ref{ts}).

\begin{figure}
\centerline{\psfig{figure=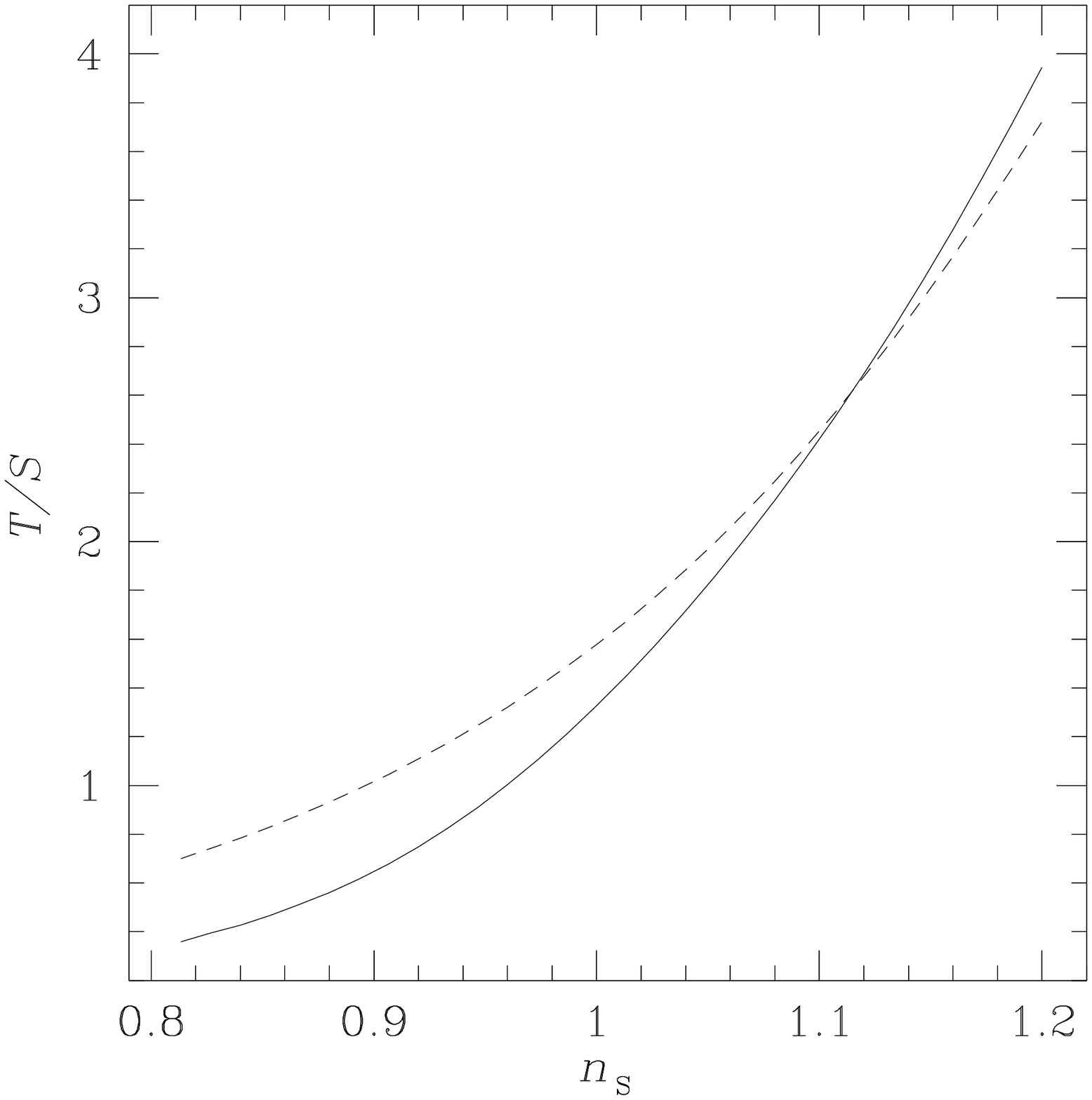,height=10cm}}
\caption{$95\%$ confidence limits on $T/S$ versus $n_{\rm S}$, for 
$T/S$ free and for binned CMB (dashed) and combined data (solid).
No parameter constraints have been applied.}
\label{bluefig}
\end{figure}

In Fig.~\ref{freefig} we see that 
the data are considerably less constraining, compared with the PLI case,
when we allow $T/S$ to vary freely.  This was expected, since 
in the PLI case, the lowering of $n_{\rm S}$ tends to enhance the effect of 
increasing $T/S$.  We also expect that a blue scalar
tilt would oppose the effect of tensors on the spectrum and
hence allow larger $T/S$.  Figure \ref{bluefig} illustrates this
effect by plotting the $95\%$ upper confidence limits on $T/S$ 
versus scalar index with $T/S$ free.  Note that we cannot 
meaningfully constrain $T/S$ here by marginalizing over $n_{\rm S}$,
since the best fits to the spectrum remain good even for very
blue tilts and very large $T/S$.

Our confidence limits are summarized in Table~\ref{limpli} for the 
case of power-law inflation, Table~\ref{limpoly} for polymomial 
potentials, and Table~\ref{limgen} for free $T/S$ and $n_{\rm S}=1$.  
In all cases $95\%$ upper limits on $T/S$ are presented, after 
integrating over the ranges of parameter space specified in 
Sec.~\ref{cmb}.  The row gives the data set used, while 
the column specifies the type of parameter constraint applied, if any.  

\begin{table}
\caption{$95\%$ confidence limits on $T/S$ for various 
data sets and parameter constraints, and for power-law inflation.
``$2.5+$'' represents no constraint (for values as high as this we do
not expect our approximations to be adequate in any case).}
\begin{tabular}{cddddd}
Data set        & No constraint & $t_0>11$ Gyr & $t_0>13$ Gyr & Cluster baryon 
                    & Supernova \\
\tableline
{\sl COBE\/}        & 2.1  & 2.1  & 2.1  & 2.1  & 2.1  \\
binned CMB          & 0.60 & 0.62 & 0.67 & 0.63 & 0.65 \\
galaxy correlations & 2.5+ & 2.5+ & 2.5+ & 2.5+ & 2.5+ \\
cluster abundance   & 1.8  & 1.4  & 1.1  & 1.5  & 1.1  \\
\lya\ absorption    & 1.6  & 1.1  & 0.82 & 1.3  & 0.96 \\
combined            & 0.52 & 0.52 & 0.51 & 0.53 & 0.47 \\
\end{tabular}
\label{limpli}
\end{table}

\begin{table}
\caption{$95\%$ confidence limits for $T/S$ 
as for Table~\ref{limpli} but for $\phi^p$ inflation.}
\begin{tabular}{cddddd}
Data set        & No constraint & $t_0>11$ Gyr & $t_0>13$ Gyr & Cluster baryon 
                    & Supernova \\
\tableline
%{\sl COBE\/}       &  &  &  &  &  \\
binned CMB          & 0.63 & 0.64 & 0.67 & 0.64 & 0.65 \\
galaxy correlations & 2.5+ & 2.5+ & 2.5+ & 2.5+ & 2.5+ \\
cluster abundance   & 2.1  & 1.5  & 1.2  & 1.8  & 1.2  \\
\lya\ absorption    & 1.8  & 1.2  & 0.87 & 1.5  & 1.1  \\
combined            & 0.49 & 0.49 & 0.49 & 0.50 & 0.45 \\
\end{tabular}
\label{limpoly}
\end{table}

\begin{table}
\caption{$95\%$ confidence limits on $T/S$ as for Table~\ref{limpli}
but for $n_{\rm S}=1$ and $T/S$ free.}
\begin{tabular}{cddddd}
Data set        & No constraint & $t_0>11$ Gyr & $t_0>13$ Gyr & Cluster baryon 
                    & Supernova \\
\tableline
{\sl COBE\/}        & 2.5+ & 2.5+ & 2.5+ & 2.5+ & 2.5+ \\
binned CMB          & 1.6  & 1.6  & 1.8  & 1.6  & 1.8  \\
cluster abundance   & 2.5+ & 2.5+ & 2.5+ & 2.5+ & 2.5+ \\
\lya\ absorption    & 2.5+ & 2.5+ & 2.5+ & 2.5+ & 2.5+ \\
combined            & 1.3  & 1.3  & 1.3  & 1.3  & 1.0  \\
\end{tabular}
\label{limgen}
\end{table}

\section{The Future}

The discovery of a nearly scale-invariant spectrum of long wavelength
gravity waves would be tremendously illuminating.
Inflation is the only known mechanism for producing an almost scale-invariant
spectrum of adiabatic scalar fluctuations, a prediction which is slowly
gaining observational support.
In the simplest, ``toy'', models of inflation a potentially large amplitude
almost scale-invariant spectrum of gravity waves is also predicted.
For monomial inflation models within the slow-roll approximation, detailed
characterization of this spectrum could in principle allow a reconstruction
of the inflaton potential \cite{InfReview}.
This surely is a window onto physics at higher energies than have ever been
probed before.

Inflation models based on particle physics, rather than ``toy'' potentials,
predict a very small tensor spectrum \cite{lyth97}.  However, essentially
nothing is known about particle physics above the electroweak scale, and
extrapolations of our current ideas to arbitrarily high energies could easily
miss the mark.  We must be guided then by observations.
We have argued that observational support for a large gravity wave component
is weak.  Indeed observations definitely require the tensor anisotropy to
be subdominant for large angle CMB anisotropies.
One the other hand, it is still possible to have $T/S\simeq0.5$, and since
it would be so exciting to discover any tensor signal at all we are led
to ask: how small can a tensor component be and still be detectable?
What are the best ways to look for a tensor signal?

\subsection{Direct detection}

The feasibility of the direct detection of inflation-produced gravitational
waves has been addressed by a number of authors
\cite{kw,White,TurWhiLid,TurWhi,lid94,t,CalKamWad},
with pessimism expressed by most.

The ground-based laser interferometers LIGO and VIRGO \cite{thorne} will
operate in the $f \sim 100\,$Hz frequency band, while the European Space
Agency's proposed space-based interferometer LISA \cite{danz} would operate
in the $f \sim 10^{-4}\,$Hz band.  Millisecond pulsar timing is sensitive to
waves with periods on the order of the observation time, \ie frequencies
$f \sim 10^{-7}-10^{-9}\,$Hz \cite{thorne}.
These instruments probe regions of the tensor perturbation spectrum which
entered during the radiation dominated era.
Expressions for the fraction of the critical density due to gravity waves
per logarithmic frequency interval can be found in
\cite{White,TurWhiLid,TurWhi,lid94,t}.
Assuming that $\om0=1$ in a PLI model, with the only relativistic particles
being photons and 3 neutrino species, and taking the {\sl COBE\/} quadrupole
$Q=T+S\simeq4.4\times10^{-11}$, one finds \cite{t}
\beq
\Omega_{{\rm GW}}(f)h^2 = 5.1\times10^{-15} \frac{n_{\rm T}}{n_{\rm T}-1/7} 
                \exp\left[n_{\rm T}N+\frac{1}{2}N^2(dn_{\rm T}/d\ln k)\right],
\label{omg}
\eeq
where $N\equiv\ln(k/H_0)$ and $n_{\rm T}=-(T/S)/7$ is the tensor spectral
index.

Using \eq(\ref{omg}) Turner \cite{t} found that the local energy density in
gravity waves is maximized at $T/S=0.18$ for $f \sim 10^{-4}\,$Hz.
At this maximum, the local energy density is in the range
$\Omega_{{\rm GW}}h^2 \simeq 10^{-15} - 10^{-16}$, which lies a couple of
orders of magnitude below the expected sensitivity of LISA, and several
orders below that of LIGO/VIRGO \cite{thorne}.
This is also well below the current upper limit of
$\Omega_{{\rm GW}}h^2 < 6\times10^{-8}$ (at $95\%$ confidence) from pulsar
timing \cite{ktr}.
As $T/S$ increases above $0.18$, $\Omega_{{\rm GW}}(f\sim10^{-4}\mbox{Hz})h^2$
begins to {\it decrease\/} due to the increasing magnitude of the tensor
spectral index.

Recall that our joint data constraint for PLI gives $T/S \alt 0.5$, so our
results predict that the inflationary spectrum of gravity waves from PLI is
not amenable to direct detection.

\subsection{Limits from the CMB}

With the advent of {\sl MAP\/} and especially the {\sl Planck\/} Surveyor, with
its higher sensitivity, detailed maps of the CMB are just around the corner.
What do we expect will be possible from these missions?
This question has been dealt with extensively before.
Assuming a cosmic variance limited experiment capable of determining
only the anisotropy in the CMB but with all other parameters known, one
can measure $T/S$ only if it is larger than about 10\% \cite{KnoTur}.
A more realistic assessment for {\sl MAP\/} and {\sl Planck\/} suggests
this limit is rarely reached in practice \cite{ZalSpeSel,BET}.

However the ability to measure linear polarization in the CMB anisotropy
offers the prospect of improving the sensitivity to tensor modes
(for a recent review of polarization see \cite{Polar}).
In addition to the temperature anisotropy, two components of the
linear polarization can be measured.  It is convenient to split the
polarization into parity even ($E$-mode) and parity odd ($B$-mode)
combinations -- named after the familiar parity transformation properties of
the electric and magnetic fields, but not to be confused with the $E$ and
$B$ fields of the electromagnetic radiation.

Polarization offers two advantages over the temperature.  First, with more
observables the error bars on parameters are tightened.  In addition the
polarization breaks the degeneracy between reionization and a tensor
component, allowing extraction of smaller levels of signal \cite{ZalSel}.
Model dependent constraints on a tensor mode as low as $1\%$ appear to
be possible with the {\it Planck\/} satellite
\cite{ZalSpeSel,BET,KamKos,Kinney}.  Extensive observations of patches
of the sky from the ground (or satellites even further into the future)
could in principle push the sensitivity even deeper.

There is a further handle on the tensor signal however.
Since scalar modes have no ``handedness'' they generate only parity even,
or $E$-mode polarization \cite{ZalSel,KKS}.  A definitive detection of
$B$-mode polarization would thus indicate the presence of other modes,
with tensors being more likely since vector modes decay cosmologically.
Moreover a comparison of the $B$-mode, $E$-mode and temperature signals
can definitively distinguish tensors from other sources of perturbation
(\eg \cite{HuWhiPRD}).

\begin{figure}
\centerline{\psfig{figure=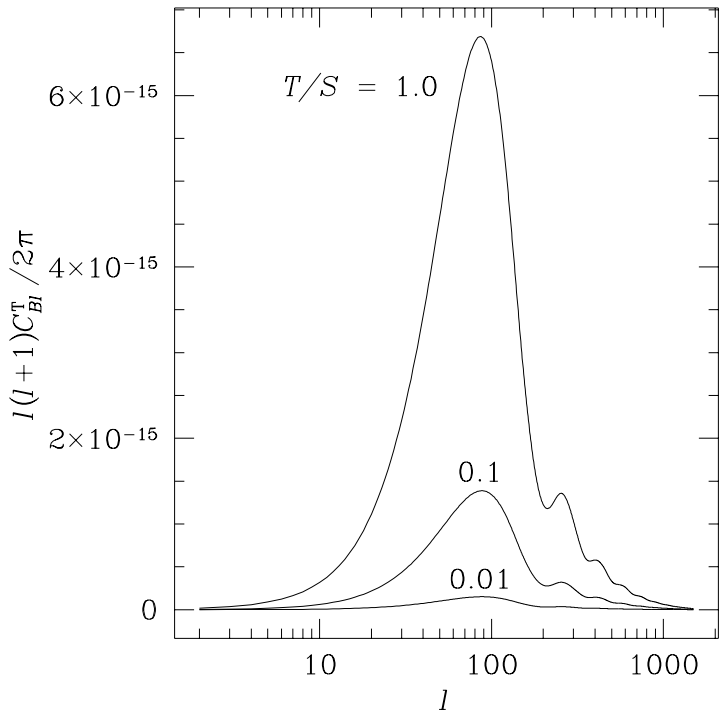,height=10cm}}
\caption{$B$-mode tensor polarization signal $C_{B\ell}^{{\rm T}}$ for 
$T/S=0.01$, 0.1, and 1.0, with the remaining parameters specified 
as standard CDM.}
\label{curlfig}
\end{figure}

Unfortunately the detection of a $B$-mode polarization will prove a
formidable experimental challenge.  The level of the signal, shown in 
Fig.~\ref{curlfig}
for $T/S=0.01$, 0.1 and 1.0, is very small.  As an indicative number, with
$T/S=0.5$, our upper limit, the {\it total\/} rms $B$-mode signal, integrated
over $\ell$, is $0.24\,\mu$K in a critical density universe.  These
sensitivity requirements, coupled with our current poor state of knowledge
of the relevant polarized foregrounds make it seem unlikely a $B$-mode signal
will be detected in the near future.

\section{Conclusions}

We have examined the current experimental limits on the tensor-to-scalar
ratio.  
Using the {\sl COBE\/} results, as well as small-scale CMB observations,
and measurements of galaxy correlations, cluster abundances, and Ly$\,\alpha$
absorption we have obtained conservative limits on the tensor fraction for some
specific inflationary models.  Importantly, we have considered models with
a wide range of cosmological parameters, rather than fixing the values
of $\Omega_0$, $H_0$, etc.
For power-law inflation, for example, we find that $T/S<0.52$ at the
$95\%$ confidence level.  Similar constraints apply to $\phi^p$ inflaton
models, corresponding to approximately $p<8$.
Much of this constraint on the tensor-to-scalar ratio comes from the relation
between $T/S$ and the scalar spectral index $n_{\rm S}$ in these theories.
For models with tensor amplitude unrelated to {\it scalar\/} spectral index
it is still possible to have $T/S>1$.
Currently the tightest constraint is provided by the combined CMB data
sets.  Since the quality of such data are expected to improve dramatically
in the near future, we expect much tighter constraints (or more
interestingly a real detection) in the coming years.

\acknowledgements
This research was supported by the Natural Sciences and Engineering 
Research Council of Canada.
M.W. is supported by the NSF.

%\bibliography{bib}

\end{document}